# PARKER'S DYNAMO AND GEOMAGNETIC REVERSALS

M. RESHETNYAK[1], D. SOKOLOFF[2]

[1]Institute of the Physics of the Earth, Russian Acad. Sci., 123995, Moscow, Russia
e-mail adress: m.reshetnyak@gmail.com
[2]Department of Physics, Moscow State University, 119899, Moscow, Russia
e-mail adress: sokoloff@dds.srcc.msu.su

In accordance with basics of the dynamo theory the observable large-scale magnetic fields in the various astrophysical bodies associate with the break of the mirror symmetry of the flow. Usually, this break is caused by rotation of the body [1]. It appears, that existence of rotation and density gradient is enough for correlation of the turbulent velocity **v** and its vorticity **w**=rot **v**, what leads to the non-zero kinematic helicity: $\chi=<\mathbf{v}\cdot\mathbf{w}>$, where <...> denotes averaging. In the simplest case one assumes, that helicity has negative sign (for the self-gravitating bodies) in the northern hemisphere and positive sign in the southern one, e.g. in the form $\chi=-\cos(\vartheta)$, where $\vartheta$ is an angle from the axis of rotation. Existence of the non-zero $\chi$ makes easier generation of the large-scale magnetic field.

However there are many reasons why $\chi$ in the realistic physical applications can have more complicated form. Some models, where together with equations for the magnetic field **B** differential equation for $\chi$ is solved [2]. Even for the geodynamo problems with prescribed $\chi$ one can expect that the non-uniform heat-flux at the core-mantle boundary can produce some fluctuations of $\chi$. Similar situation can arise for the binary connected objects. Moreover, these fluctuations can arise as a result of the limited number of statistics when $\chi$ is computed for the particular velocity field **v** [3]. These ideas motivated us to study how already for the kinematic regime the small fluctuations of helicty $\chi$ (and related with it $\alpha$-effect) can change the final form of the linear solution. For this aim we considered the eigen value problem for the well-known Parker's dynamo equations. The obtained results are relevant for the geodynamo and solar dynamo applications. We also consider important point what sign of the dynamo number in the Parker's dynamo is more suitable for the geodynamo applications and its relation with reversals of the geomagnetic field.

One of the simplest dynamo models is a one-dimensional Pakrers's model of the form [1,4]:

$$\partial A/\partial t=\alpha B+A'', \quad \partial B/\partial t=-D\sin(\vartheta)A'+B'',$$

where A and B are azimuthal components of the vector potential and magnetic field, *D* is a dynamo number, which is a product of the amplitudes of the $\alpha$- and $\omega$-effects and primes denote derivatives with respect to $\vartheta$. Equations are solved in the interval $0\leq\vartheta\leq\pi$ with the boundary conditions A=B=0 at $\vartheta=0, \pi$. Generation of the magnetic field ($B_r$, B), where $B_r$=A' - is a radial component of the magnetic field, is a threshold phenomenon. It starts, when *D* is larger than some critical value $D_c$. Here we consider the eigen value problem for these equations with $\alpha=\cos(\vartheta)$, see Fig.1a,b, where the complex growth rate $\gamma$ for the first three modes is presented. The positive values of the real part $\Re\gamma$ correspond to the growing solutions. These modes are sorted in such a way, that the maximal growth rate is denoted with circles, then follow squares, and the minimal growth rate corresponds to triangles. With increase of *D* (*D*>0) the first stationary (S) mode (which imaginary part $\Im\gamma$ is zero) appears. This mode is a quadrupole (Q). Increase of *D* leads to the bifurcation of the second oscillatory dipole mode (OD). $\Im\gamma$ characterizes frequency of oscillations. Symmetry of $\Im\gamma$ in respect to *D*-axis is connected with existence of the conjugate solution. It appears that with increase of *D* $\Re\gamma_2$ becomes larger than $\Re\gamma_1$. The oscillatory solution is a wave, which propagates from the equator to the poles.

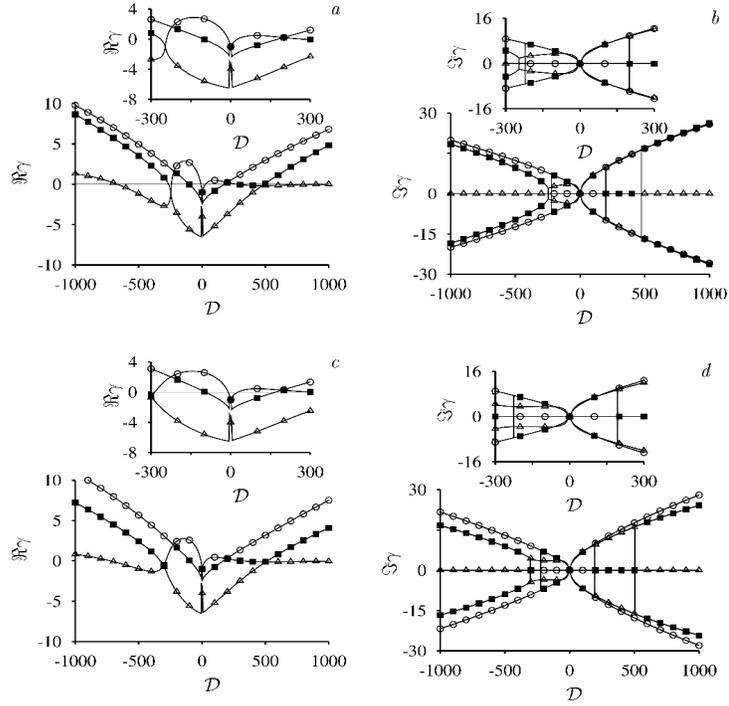

**Figure 1.** Dependence of $\Re\gamma$ (*a*) and $\Im\gamma$ (*b*) on *D* for *C*=0 (*a, b*) and *C*=0.1 (*c, d*) for the first three modes in the order of decaying $\Re\gamma$ (circles, squares and triangles). The insets correspond to the scaled-up regions.

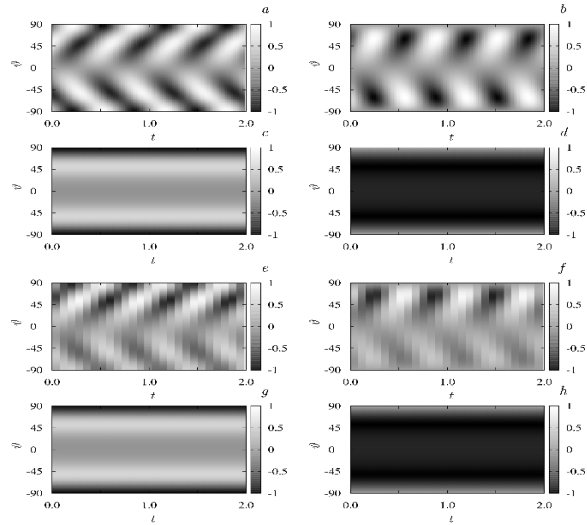

**Figure 2.** The butterfly diagrams of $B_r$ (left column) and B (right column) of the oscillatory dipole (*a, b*) and stationary quadrupole solutions (*c, d*) for *D*=200 and *C*=0. Similar plots *e, f, g, h* for *C*=0.1. The dipole solution corresponds to the circles, quadrupole solution - to the squares, see Fig.1. The typical blurring for the oscillatory solution is observed. The stationary (quadrupole) solution has a weak dependence on *C*.

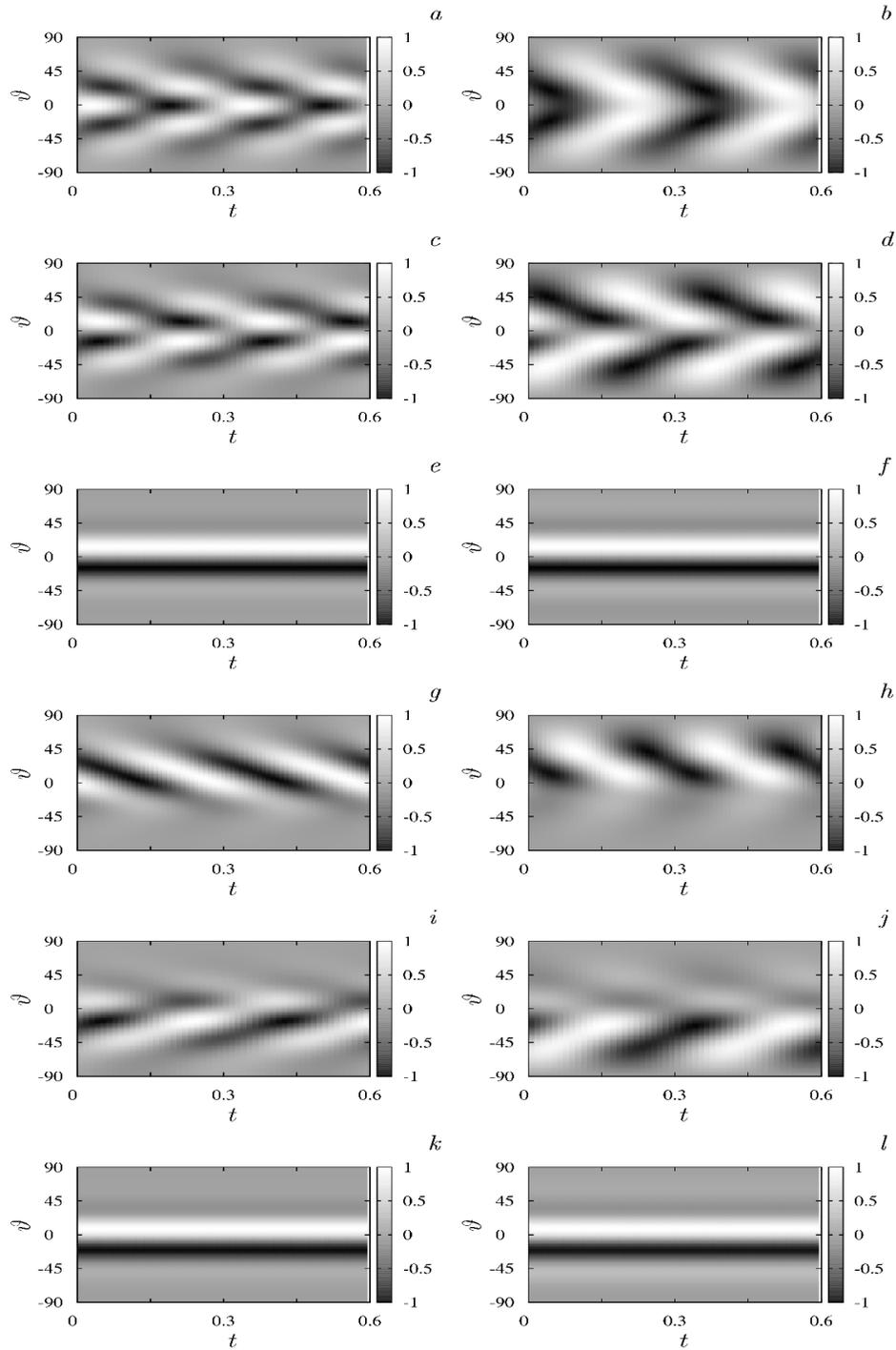

**Figure 3.** The butterfly diagrams of $B_r$ (left column) and $B$ (right column) of the oscillatory quadrupole (*a, b*), oscillatory dipole (*c, d*) and stationary dipole (*e, f*) solutions for $D=-1000$ and $C=0$. Similar plots *g, h, i, j, k, l* for $C=0.1$. The quadrupole solution corresponds to the circles, and two dipole solution - to the squares and triangles, see Fig.1. Non-zero $C$ leads to i) penetration of $B_r$-quadrupole wave from the northern hemisphere to the southern one and damping of the quadrupole B-wave in the southern hemisphere; ii) damping of the oscillatory dipole $B_r$- and B-waves in the northern hemisphere. The stationary solution is still unchanged.

The further increase of *D* leads the change of SQ-mode to OQ-mode, also propagating to the poles.

For *D*<0 the first mode is SD. Then OQ-mode, which is a wave, propagating from the poles to the equator, appears. For the large negative *D* there are already three modes: OD, OQ and SD.

To introduce fluctuation of the α-effect we consider change of the mean (zero) value of α up to some small value *C*: α→α+*C*. The first glance at the behavior of γ does not reveal any significant changes in the spectrum, see Fig.1. However it does not mean, that nothing changed in the system.

It appears, that the form of the eigen solution is more sensitive to fluctuation *C*. To plot the complex eigen functions we introduce the butterfly diagrams $\Re((B_r, B) \exp(i\Im\gamma))$, see Fig.2,3. Comparison of oscillatory modes with *C*=0 and *C*=0.1 gives that already small fluctuation of α leads to the significant change of the eigen solution: the form of solution becomes blurred Fig.2a,e and what is more important - asymmetry of the hemispheres becomes apparent. It is well observed for OD-mode for B component, see Fig.2b,f, 3d,j where the magnetic field in the southern hemisphere is smaller then in the northern one. The other interesting fact is a penetration of the $B_r$-wave through the equator plane, see Fig.3g.

To demonstrate suppression of the field in the hemispheres we present dependence of the maximal values of B for *C*=0 and *C*=0.1, see Fig.4. For the oscillatory (with zero mean value) B, $B_{max}(\theta)$ for *C*=0 is a positive symmetrical in respect to equator plane function. The non-zero *C* leads to suppression of B in the southern hemisphere. As we already noted, fluctuation *C* does not change stationary solution. For *D*<0 the non-zero *C* leads to damping of OQ-mode in the southern hemisphere and of OD-mode in the northern hemisphere.

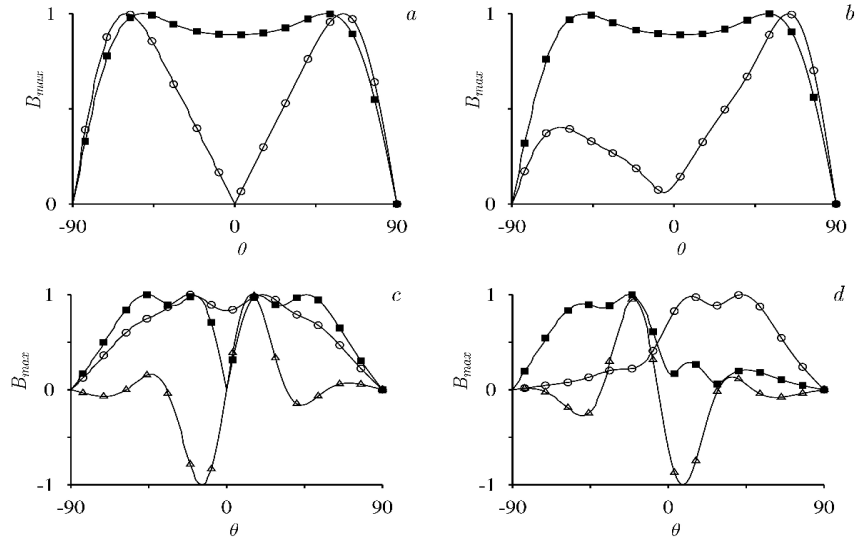

**Figure 4. The maximal values of B of the first three modes as a function of the latitude θ for D=200 (upper line) and D=-1000 (lower line) for C=0 (left column) and C=0.1 (right column). The labels are the same to that ones in Fig.1, 2, 3.**

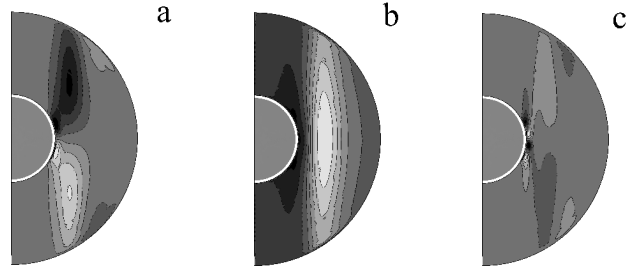

**Figure 5.** Distribution of the averaged on azimuthal direction kinematic helicity $\chi$ (a), differential rotation $\Omega$ (b) and their product $\chi\Omega$ (c) in the 3D-model of the thermal convection in the spherical shell for the modified Rayleigh number Ra=$10^2$, Prandtl number Pr=1 and Ekman number E=$10^{-4}$. The maximal values correspond to the white color, the dotted isolines - to the negative values.

The exact answer why the small value of *C* leads to suppression of generation in one hemisphere is addressed to the future linear analysis. Now we only can propose some suggestions on this problem. So as this phenomenon appears when solution is oscillatory, then we suppose, that it is concerned with the interference of the original wave and a new one caused with bifurcation *C*. It is worthy to note, that *C* has the quadrupole symmetry in respect to the equator plane. It means, that the new mode produced by *C* has opposite symmetry to the original wave. Superposition of these modes can lead to suppression of the field generation in one hemisphere.

To apply our results to the geodynamo we need to fix a sign of the dynamo number *D*, which is responsible for the direction of the wave propagation. For the solar dynamo situation is clear: the wave propagates from the poles to the equator and *D* is negative. For the geodynamo situation is more difficult. May be the first estimate of sign of *D* in the Earth's core belongs to Olson, who, based on the structure of the flows in the liquid core of the Earth supposed, that *D* is positive [5]. The more certain information on sign of *D* can be obtained from 3D numerical simulations [6]. Here we present the typical distribution of the averaged helicity $\chi$, differential rotation $\Omega=\partial V_\varphi/\partial r$ based on the radial gradient of the azimuthal velocity and their product, which give the sign of *D* (in the northern hemisphere), see Fig.4. It should be mentioned, that generally used in geodynamo Boussinesq models of convection have limited sources of helicty generation. So as liquid metal supposed to be incompressible, helicty can be generated in Ekman layes and by the cyclone convection. Increase of Rayleigh number can destroy vertical columns and move helicty to the boundary layers. Presented results in Fig.4 correspond to the Rayleigh a few times larger than its critical value. Finally, we are close to conclude, that *D* is positive in the Earth's core.

Consider the first possibility, *D*>0. Following linear analysis presented above we have transitions with increase of *D* of the form: SQ → OD+SQ → OD+OQ. Here and below modes in sum are ordered in decreasing order of $\Re\gamma$. To describe change of regime with rapid to seldom reversals we need to find such a *D*, that OD changes either to SD or to OD+SD. The latter one corresponds to the regime in vascillation. We do not observe such kind of transitions for *D*>0.

Case *D*<0, with SD → SD+OQ → OQ+OD+SD, which is more attractable for the solar dynamo, appears to be more useful for the geodynamo also. It means, that for *D* near to the threshold of generation reversals are absent. Increase of *D* leads to the bifurcation of OQ-mode, which can be associated with excursions of the geomagnetic field: the so-called failed reversals, when the virtual dipole comes to the equator plane and then returns back. The further increase of *D* gives us a new bifurcation with a dipole solution in vascillations. These oscillations accompanied with a OQ-mode which introduce asymmetry in the hemispheres.